# Some Features Of The Flow In The Holweck Pump

P. A. Skovorodko

*Institute of Thermophysics, 630090, Novosibirsk, Russia*

## ABSTRACT

Numerical algorithm for direct simulation of the gas flow in the Holweck pump is developed. The results illustrating the important features of the flow in the pump are reported. An attention is paid to the problem of the pump design optimization.

## Algorithm And Results

This study is a continuation of Ref. 1 where the flow in the Holweck pump was investigated under free molecular conditions. To take into account the effect of molecular collisions the direct simulation of the flow in the frames of the usual DSMC approach [2] is applied. Unlike the approach of Ref. 3, where the flow is treated in the rotating reference system, we use the laboratory reference system with stationary stator. The trajectories of molecules between collisions represent the straight lines while the rotation of the domain of simulation is taken into account. Two molecular models - solid spheres and Maxwell molecules (VSS) [2] with the parameters derived from the viscosity were applied, though no noticeable effect of the model of molecules on the results were found. The simulations were performed for the pump used in Ref. 4 for theoretical and experimental investigations. The results obtained in test calculations for collisionless flow were found to be in good agreement with those previously obtained by test particle Monte Carlo method [1].

Fig. 1 illustrates the distribution of pressure averaged over radial coordinate in the flow field obtained with (a) and without (b) taking into account the effect of molecular collisions. The results correspond to the helium flow with inlet pressure 0.1 Torr, back pressure 1 Torr and rotational speed of rotor $N = 400$ rps. Only a part of the total surface of the rotor corresponding to the flow field period (60 degrees [4]) is shown. The bottom of the pictures corresponds to the pump inlet, the groove channel origin being placed on the right corner. Due to the symmetry conditions the parameters on the left and right sides of the pictures are identical. The isobars are shown through 0.02 Torr.

As can be seen from Fig. 1 on significant part of the flow field the isobars are perpendicular to the inlet surface but not to the back surfaces of the groove channel. This peculiarity is typical of the flow in the screw-grooved pump and reflects an important feature of the flow from the point of view of its approximate models. The effect of molecular collisions in the considered conditions is mainly pronounced for the inlet part of the pump, the gas flow rate being reduced by about 20% compared to the free molecular flow.

Concerning the problem of the pump design optimization it should be noted that a large number of parameters determines the pump performance. Since the parameters of "optimum" pump depend on many factors such as working gas, the pressure range, etc., the process of the pump design optimization represents the finding of some compromise solution of the problem. In this study the effect of two determining parameters on the pump operation is analyzed: the number of the groove channels on the rotor surface ($n$) and the helix angle ($\alpha$) between the direction of the groove channel and the inlet surface. These parameters were varied in the range $n \geq 2$, $10° \leq \alpha \leq 60°$, while in the original pump $n = 6$, $\alpha = 25°$ [4]. The study was performed for air as a working gas and $N = 400$ rps. To check the idea that the parameters of "optimum" pump operating in the free molecular conditions will be close to those in a real flow on the preliminary stage of the study the optimization was performed for collisionless flow. The obtained results are as follows: the optimum value of $\alpha$ for maximum throughput is close to $45°$ and it doesn't depend significantly on $n$. The optimum value of $\alpha$ for maximum compression ratio depends on $n$ and is about $25°$ for $n=6$ and about $30°$ for $n=12$. The optimum value of $n$ for maximum compression ratio is close to 12. Therefore, the pump with the parameters $n \approx 12$ and $\alpha = 30° \div 35°$ seemed to be more optimum than the original one [4].

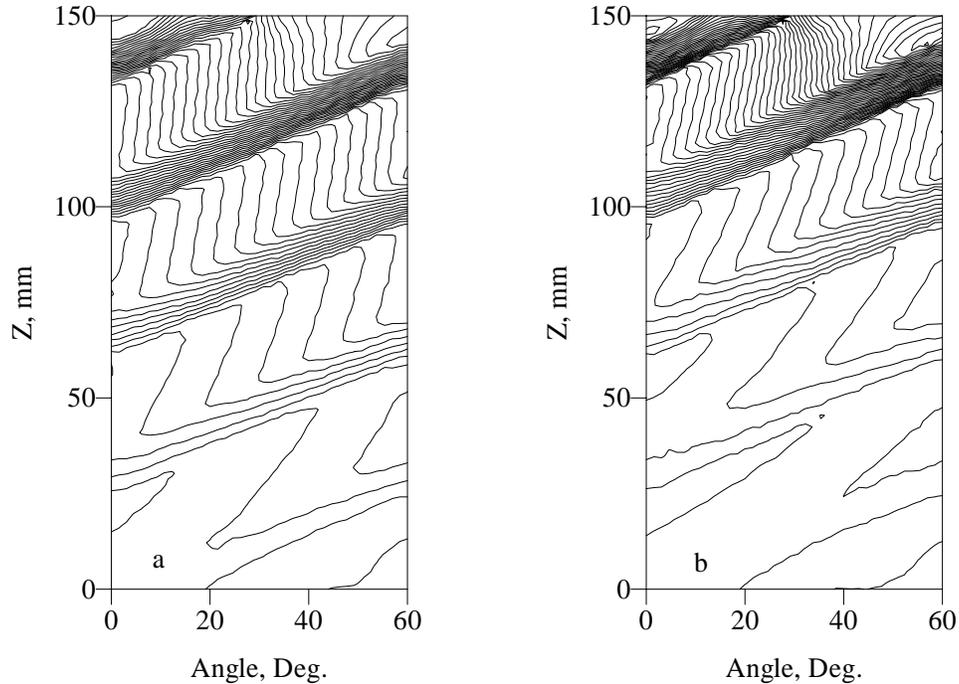

**FIGURE 1.** Radially averaged pressure distribution in the flow field obtained with (a) and without (b) taking into account the effect of molecular collisions.

## Conclusion

Though the direct simulation of the flow in the Holweck pump allows one to predict quite adequite results the possibilities of this approach to the pump design optimization seemed to be limited due to high requirements of computational resources and a large number of the parameters determining the pump performance. The approach based on the optimization of the pump operating in the collisionless flow seemed to be perspective. The development of approximate models of the flow [4,5] is still quite actual too. Such models should be based on the real properties of the flow, that may be just found by DSMC approach. One of the features of the flow in the Holweck pump is that on significant part of the flow field the isobars are perpendicular to the inlet surface but not to the back surfaces of the groove channel.

## REFERENCES

1. Skovorodko, P., A., "Free Molecular Flow in the Holweck Pump", in Rarefied Gas Dynamics-2000, edited by T. J. Bartel and M. A. Gallis, 22nd International Symposium Proceedings, AIP Conference Proceedings, 2001, Vol. 585, pp. 900 - 902.
2. Bird, G. A., Molecular Gas Dynamics and the Direct Simulation of Gas Flows, Clarendon Press, Oxford, 1994.
3. Hwang, Y.-K., and Heo, J.-S., "Molecular Transition and Slip Flows in Rotating Helical Channels of Drag Pump", in Rarefied Gas Dynamics-2000, edited by T. J. Bartel and M. A. Gallis, 22nd International Symposium Proceedings, AIP Conference Proceedings, 2001, Vol. 585, pp. 893-899.
4. Kanki, T., "Flow of a Rarefied Gas in a Rectangular Channel with a Moving Plate", in Rarefied Gas Dynamics-1994, edited by J. Harvey and G. Lord, 19th International Symposium Proceedings, Oxford University Press, Oxford, 1995, Vol. 1, pp. 375 - 381.
5. Spagnol, M., Cerruti, R., and Helmer, J., J. Vac. Sci. Technol. A 16(3), 1151-1156 (1998).